\documentclass{article}

\usepackage{amsmath}

\usepackage{amssymb}

\usepackage{cite}

\title{Type V singularities with inhomogeneous equations of state }

\author{Oem Trivedi \footnote{oem.t@ahduni.edu.in} \\ School of Arts and Sciences \\ International Centre for Space and Cosmology \\   Ahmedabad University \\  Navrangpura, Ahmedabad, 380009  \\ Gujarat,India}

\begin{document}
	
	\maketitle
	
	\begin{abstract}
		Interest in cosmological singularities has remarkably grown in recent times, particularly on future singularities with the discovery of late-time acceleration of the universe and dark energy. Recent work has seen a proper classification of such singularities into strong and weak based on their strength, with weak singularities being the likes of sudden, w and big freeze singularities and strong singularities like the big rip. This has led to a classification of such singularities in various types like Big rip is Type 1, w-singularity is type V etc. While singularities of type I-type IV have been discussed vividly by taking into account inhomogeneous equations of state (EOS), the same has not been attempted for type V singularities. So in this work we have discussed in detail about the formation of type V singularities in various cosmologies after considering inhomogeneous equations of state. We consider two inhomogeneous forms of the EOS in the context of four different cosmological backgrounds ; standard general relativistic cosmology, an asymptotically safe cosmology, a cosmology inspired by modified area-entropy relations, generalized uncertainty principles, holographic renormalization and Chern-Simons gravity( all of which can be coincidentally described by the same form of the modified Friedmann equation) and an f(R) gravity cosmology. We show in detail that one sees some very big differences in the occurence conditions of type V singularities when one makes such considerations. In the particular case of the f(R) gravity cosmology, we see that the type V singularities get completely removed. This work goes to show that the creation and formation of type V singularities is influenced most strongly by the form of the equation of state that one considers, way more so than what background cosmology one chooses.  
	\end{abstract}
	
	\section{Introduction}

	Observations of late time acceleration of the Universe came as a huge surprise to the cosmological community \cite{SupernovaSearchTeam:1998fmf} and ever since then a lot of work has been done in order to explain this expansion. The cosmological expansion problem has been addressed from multiple facets till now, which include the standard approaches of the Cosmological constant \cite{Weinberg:1988cp,Lombriser:2019jia,Padmanabhan:2002ji} alongside more exotic scenarios like Modified gravity theories\cite{Capozziello:2011et,Nojiri:2010wj,Nojiri:2017ncd} and even scalar field driven late-time cosmic acceleration scenarios \cite{Zlatev:1998tr,Tsujikawa:2013fta,Faraoni:2000wk,Gasperini:2001pc,Capozziello:2003tk,Capozziello:2002rd}. Another very interesting way to address cosmological concerns which has picked up pace in recent years is a Renormalization group approach to cosmology, leading to the so called asymptotically safe cosmology.
	\\
	\\
	There has also been an expansive literature in recent times which has been devoted to the study of various types of singularities that could occur during the current and far future of the Universe, with the observation of late-time acceleration having given a significant boost to such works \cite{Nojiri:2004ip,Nojiri:2005sr,Nojiri:2005sx,Bamba:2008ut,Trivedi:2022ipa,Trivedi:2022ngt,odintsov2015singular,odintsov2016singular,oikonomou2015singular,nojiri2015singular,Odintsov:2022eqm,Nojiri:2004pf,Capozziello:2005pa,Bamba:2012cp,Odintsov:2021yva,Nojiri:2022xdo,Brevik:2021wzs}. These cosmological singularities which have been discussed in recent times can be classified broadly into two types ; strong and weak. Strong singularities are those singularities which can distort finite objects and can mark either the beginning or the end of the universe, with the big bang being the one for the start of the universe and the so called "big rip" signalling the end of the universe. Weak singularities, as the name might suggest, are those which do not have such far reaching implications and do not distort finite objects in the same sense as their strong counterparts. We can discuss these various singularities in more detail as follows : \begin{itemize}
		\item Type I ("Big Rip") : In this case, the scale factor , effective energy density and effective pressure density diverges. This is a scenario of universal death, wherein everything which resides in the universe is progressively torn apart \cite{Caldwell:2003vq}.
		\item Type II ("Sudden/Quiescent singularity") : In this case, the pressure density diverges and so does the derivatives of the scalar factor from the second derivative onwards \cite{Barrow:2004xh}. The weak and strong energy conditions hold for this singularity. Also known as quiescent singularities, but this name originally appeared in contexts related to non-oscillatory singularities \cite{Andersson:2000cv}. A special case of this is the big brake singularity \cite{Gorini:2003wa}.
		\item Type III ("Big Freeze") : In this case, the derivative of the scale factor from the first derivative on wards diverges. These were detected in generalized Chaplygin gas models \cite{bouhmadi2008worse}.
		\item Type IV ("Generalized sudden singularities"): These are finite time singularities with finite density and pressure instead of diverging pressure. In this case, the derivative of the scale factor diverges from a derivative higher than the second \cite{Bamba:2008ut}.
		\item Type V ("w-singularities") : In this case, the scale factor, the energy and pressure densities are all finite but the barotropic index $w = \frac{p}{\rho}$ becomes singular \cite{Fernandez-Jambrina:2010ngm}. 
		\item Type $\infty$(" Directional singularities "): Curvature scalars vanish at the singularity but there are causal geodesics along which the curvature components diverge \cite{Fernandez-Jambrina:2007ohv} and in this sense, the singularity is encountered just for some observers. 
		\item Inaccessible singularities: These singularities appear in cosmological models with toral spatial sections, due to infinite winding of trajectories around the
		tori. For instance, compactifying spatial sections of the de Sitter model to cubic tori. However, these singularities cannot be reached by physically well defined observers and hence this prompts the name inaccessible singularities \cite{mcinnes2007inaccessible}.
	\end{itemize}
	In recent times, there has been a wide interest in dark energy models based in exotic non general relativistic regimes particularly because such theories display properties which are not evident in conventional cosmological models. For example,a lot of works have considered the possibility of viable scalar field based dark energy regimes in quantum gravity corrected cosmologies like the RS-II Braneworld and Loop Quantum Cosmology \cite{Sahni:2002dx,Sami:2004xk,Tretyakov:2005en,Chen:2008ca,Fu:2008gh}. There has been substantial work on new dark energy models based on thermodynamic modifications like modified area-entropy relations as well\cite{Tavayef:2018xwx,Radicella:2011qpl,Bamba:2009id,Younas:2018kmy,Jawad:2016tne,Nojiri:2019skr}. A great amount of literature has also been devoted to look out for exotic dark energy regimes based in cosmologies where the generalized uncertainty principles \cite{maggiore1994quantum,Adler:2001vs,tawfik2014generalized,Barca:2021epy} are considered instead of the usual Heisenberg uncertainty criterion \cite{Ghosh:2011ft,Rashki:2019mde,Paliathanasis:2021egx}. There has been a lot of work devoted towards studying dark energy regimes in non-canonical approaches like DBI etc. as well \cite{Calcagni:2006ge,Gumjudpai:2009uy,Chiba:2009nh,Ahn:2009xd,Li:2016grl,mandal2021dynamical,Kar:2021gbz}.
	\\
	\\
	This vast dark energy literature has prompted the study of cosmological singularities in a wide range of cosmological backgrounds as well, as there have been multiple works which have discussed Type I-IV singularities in various cosmologies \cite{Shtanov:2002ek,Bamba:2012ka,Bamba:2010wfw,Nojiri:2008fk,odintsov2018dynamical,Odintsov:2018awm,Trivedi:2022ipa,Bombacigno:2021bpk,Nojiri:2006gh,Fernandez-Jambrina:2021foi,Chimento:2015gga,Chimento:2015gum,Cataldo:2017nck,Nojiri:2005sr,Nojiri:2005sx,trivedi2022singularity}. Recently, type V singularities have again started to gain attention \cite{Trivedi:2022ngt,Ozulker:2022slu} but still, type V singularities have not been investigated in as detailed a fashion as singularities of the type I- type IV. At this point, a discussion is in order over the cosmological significance of w-singularities. While Type I-Type IV singularities deal with more direct cosmological parameters like the scale factor, Hubble parameter alongside energy and pressure densities, type V singularities deal with a somewhat indirect parameter in the form of w. This is not say, however, that these singularities cannot occur in cosmological and in particular, dark energy models. For example, it was very interestingly shown in \cite{Ozulker:2022slu} that in some models which are aimed at solving arguably the richest and most heavily debated observational tension today, the Hubble tension \cite{DiValentino:2021izs}, one could have quite viable w-singularities whilst still retaining the qualities of these models which help them address the H0 tension. In \cite{Elizalde:2018ahd} it was discussed how w-singularities can occur in interacting dark energy models(while the background cosmology in this case was still general relativistic and the continuity equation had its usual form), while in \cite{Khurshudyan:2018kfk} it was showed how varying Chaplygin gas models can also have w-singularities. Furthermore another observational clue towards w-singularities was given in this paper in which they predicted that one could expect such singularities to occur at high redshifts. It has also been discussed that w-singularities can take place alongside big brake singularities when one considers, for example, observationally viable tachyonic models \cite{Yurov:2009gj,Keresztes:2009vc}. In \cite{Bahamonde:2021gfp} it was shown that these singularities can occur in teleparellel theoreis of gravity as well. The occurrence of w-singularities in various other contexts has also been discussed in \cite{Szydlowski:2017evb,Samanta:2017qnn,Sadri:2018lzz,Astashenok:2012tv}.  Hence while type V singularities deal primarily with a more indirect cosmological parameter, it by no means diminishes its cosmological importance and it does appear in a variety of cosmological expansion scenarios. This encourages one towards an endeavour which tries to understand the status quo of w-singularities in non-standard cosmologies, which we will be undertaking here. For a detailed account of how these singularities can occur in different ways, see \cite{Fernandez-Jambrina:2010ngm}. Furthermore, Type I-Type IV singularities were also studied in the context of standard cosmology with inhomogeneous equations of state (EOS), firstly in \cite{Nojiri:2005sr,Nojiri:2005sx}. The authors of these papers discussed singularity structures in a cosmology with multiple types of inhomogeneous EOS and showed that Type I-IV singularities can occur in such a cosmology if certain conditions are fulfilled. They primarily considered an EOS of the form \begin{equation}
	p = - \rho - f(\rho)
	\end{equation} p and $\rho$ correspond to pressure and energy densities, respectively.  They also showed how there can be a transition between phantom and quintessence types of evolution for the universes given various inhomogeneous forms for the EOS. Similar endeavour in the context of an asymptotically safe cosmology was undertaken in \cite{trivedi2022singularity}. In a recent paper \cite{Trivedi:2022ngt}, it was shown that w-singularities occur in almost the same conditions in a variety of exotic cosmological backgrounds as they do in standard cosmology. So in this work we would like to see whether considering inhomogenous equations of state could influence the occurrence conditions of w-singularities in a cosmology. In section II we would give a brief overview of the cosmologies we have considered while in section III we would work out these singularities with an exponential ansatz for the scale factor and finally in section IV we summarize our work and give some concluding comments.
	\section{Brief review of the cosmologies considered}  
	We would be considering four cosmological backgrounds ; General relativistic cosmology, Chern-Simons cosmology, an f(R) gravity cosmology and an asymtotically safe cosmology. For obvious reasons, we won't be giving an extended overview of GR here as that is the standard cosmological model. The $F(R)$ gravity scenario that we would like to consider has the action \cite{Carroll2004} \begin{equation}
	S = \frac{m_{p}^2}{2} \int d^4 x \sqrt{-g} \left(R - \frac{\alpha^2}{R}\right) + \int d^4 x \sqrt{-g} \mathcal{L}_{m}
	\end{equation}
	where $\alpha$ is a constant which has the units of mass, $\mathcal{L}_{m} $ is the Lagrangian density for matter and $m_{p}$ is the reduced planck's constant. The field equation for this action is 
	\begin{multline}
	\left(1 + \frac{\alpha^2}{R^2}\right) R_{\mu \nu} - \frac{1}{2} \left(1 - \frac{\alpha^2}{R^2}\right) R g_{\mu \nu} + \\  \alpha^2 \left[g_{\mu \nu} \nabla_{a} \nabla^{a} - \nabla (_{\mu} \nabla_{\nu})  \right] R^{-2} =  \frac{T_{\mu \nu}^{M}}{m_{p}^2}
	\end{multline}
	where $T_{\mu \nu}^{M}$ is the matter energy-momentum tensor. The Friedmann equation in this case can take the form \begin{equation}
	\frac{6 H^2 - \frac{\alpha}{2}}{11/8 - \frac{8 H^2}{4 \alpha}} = \frac{\rho}{3}
	\end{equation} where $\rho$ is the total energy density. This $F(R)$ gravity regime was used to explain late time cosmic acceleration as an alternative to dark energy in \cite{Carroll2004}. The action prompts one towards the notion that very tiny corrections to the usual Einstein Hilbert in the form of $R^{n}$ with $n<0$ can produce cosmic acceleration. As corrections of the form $R^n$ with $n>0$ can lead to inflation in the early universe \cite{Starobinsky:1980te}, the authors in \cite{Carroll2004} proposed a purely garvitational paradigm through (5) to explain both the early and late time accelerations of the universe. Explaining the current epoch of the universe through such a modified gravity model would in principle eliminate the need of dark energy and hence, its an interesting scenario to consider w-singularities in as well.
	\\
	\\
	The third modified Friedmann equation that we will be covering is given by \begin{equation}
	H^{2} - \alpha H^{4} = \frac{\rho}{3}
	\end{equation}
	The above equation is very special in the sense that it can be derived from various distinct approaches. This equation can be achieved by considering a quantum corrected entropy-area relation of $ S = \frac{A}{4} - \alpha \ln \left( \frac{A}{4} \right) $ where A is the area of the apparent horizon and $\alpha$ is a dimensionless positive constant determined by the conformal anomaly of the fields, where the conformal anomaly is interpreted as a correction to the entropy of the apparent horizon \cite{cai2008corrected} . Also, this could be derived in terms of spacetime thermodynamics together with a generalized uncertainly principle of quantum gravity\cite{lidsey2013holographic}. This Friedmann equation can also be derived by considering an anti-de Sitter-Schwarzschild black hole via holographic renormalization with appropriate boundary conditions \cite{apostolopoulos2009cosmology}. Finally, a Chern-Simons type of theory can also yield this Friedmann equation \cite{gomez2011standard}. Hence the equation (7) can derived by a wide range of approaches to gravitational physics and can be representative of the effects of these different theories on the cosmological dynamics.
	\\
	\\
	Testing Asymptotic Safety at the conceptual level requires the ability to construct approximations of the gravitational RG flow beyond the realm of perturbation theory. A very powerful framework for carrying out such computations
	is the functional renormalization group equation (FRGE) for the gravitational
	effective average action $ \Gamma_{k} $ \begin{equation}
	\partial_{k} \Gamma_{k} [g,\overline{g}] = \frac{1}{2} Tr \left[ (\Gamma_{k}^{(2)} + \mathcal{R}_{k} )^{-1} \partial_{k} \mathcal{R}_{k} \right]
	\end{equation} 
	Arguably the simplest approximation of the gravitational RG flow is obtained from projecting the FRGE onto the Einstein-Hilbert action approximating $ \Gamma_{k}$ by \cite{Bonanno:2017pkg} 
	\begin{multline}
	\Gamma_{k} = \frac{1}{16 \pi G_{k}} \int d^4 x \sqrt{-g} \left[-R + 2 \Lambda_{k}\right] + \\ \text{gauge-fixing and ghost terms}
	\end{multline}
	where R, $ \Gamma_{k} $ and $ G_{k} $ are the Ricci Scalar, the running cosmological constant and the running Newton's gravitational constant. The scale-dependence of these couplings is conveniently expressed in terms of their dimensionless counterparts as \begin{equation}
	\Lambda_{k} = k^2 \lambda_{*}
	\end{equation}
	\begin{equation}
	G_{k} = g_{*}/k^{2}
	\end{equation}
	where $g_{*} = 0.707$ and $ \lambda_{*} = 0.193$ .
	Regarding the choice of the cutoff scale k, there are a significant number of cutoff identifications to choose from \cite{Bonanno:2017pkg} . To proceed further, we consider the background metric to be that of a flat FLRW cosmology \begin{equation}
	ds^2 = -dt^2 + {a(t)}^2 (dx^2 + dy^2 + dz^2 )
	\end{equation}
	Considering a perfect fluid form for the stress-energy tensor , $ t_{\mu}^{\nu} = \text{diag} [-\rho,p,p,p] $ , one can get the Friedmann equation and the continuity equation in this scenario to be \begin{equation}
	H^{2} = \frac{8 \pi G_{k}}{3} + \frac{\Lambda_{k}}{3}
	\end{equation} 
	\begin{equation}
	\dot{\rho} + 3 H(\rho + p) = - \frac{\dot{\Lambda_{k}} + 8 \pi \dot{G_{k}}}{8 \pi G}
	\end{equation}
	The continuity equation comes from the Bianchi identity satisfied by Einstein's equations $ D^{\mu} [\lambda(t) g_{\mu \nu} - 8 \pi G(t) T_{\mu \nu} ] = 0 $ . To proceed further, we need a form for the cut-off scale k. A popular form for k where one uses k to be proportional to the Hubble parameter and so here we take k as \footnote{One might be tempted to think that this cut-off identification is arbitrary but it is hardly so.The cut-off choice can indeed be very different from this as well Taking the cut-off scale proportional to the Hubble parameter has been a recurring theme in asymptotically safe cosmology studies and one can see for example \cite{Bonanno:2017pkg} , \cite{mandal2020cosmology} , \cite{Bonanno:2011yx} to look at discussions about this choice of scale. In particular one can read the discussion in \cite{Bonanno:2017pkg}, for instance, which discusses four different types of cut-off identifications with the one where the cut-off choice is proportional to the Hubble parameter being one of them. This choice was firstly introduced in \cite{Bonanno:2007wg} and there has been significant work on this basis ever since.} \begin{equation}
	k^2 = \epsilon {H(t)}^2 
	\end{equation}
	where $\epsilon$ is an a priori undetermined positive parameter of $\mathcal{O}(1)$. Using this cutoff scale and the values (3-4), we can write the Friedmann as \begin{equation}
	H^4 \approx \frac{18 \rho}{3 \epsilon - 0.193 \epsilon^2 } = N \rho
	\end{equation}. 
	with $ N = \frac{18}{3 \epsilon - 0.193 \epsilon^2 } $. With this, we have completed a short review of the cosmologies we would be concerned with and their corresponding Friedmann equations. We will now proceed further to analyse the formation of type V singularities firstly with an exponential ansatz for the scale factor. 
	\\
	\\
	\section{Analysing type V singularities }
	In order to get viable expressions for the w-parameter and hence to eventually analyse type V singularities, we need to have a form for the scale factor. One suchansatz for the scale factor which can give w-singularities was proposed by Dabrowski and Marosek in \cite{Dabrowski:2012eb}, which has an exponential form. That ansatz is given by \begin{equation}
	a(t) = a_{s} \left( \frac{t}{t_{s}}\right)^m \exp \left(1 - \frac{t}{t_{s}}\right)^{n}
	\end{equation}
	where $a_{s}$ has the units of length and is a constant while m and n are also constants \footnote{While the ansatz on the surface looks quite different from a power series one which we will consider later on, it can be a sub case of a series ansatz within certain limits as well } . The scale factor is zero (a=0) at t=0, thus signifying the big bang singularity. One can write the first and second derivatives of the scale factor as \begin{equation}
	\dot{a}(t) = a(t) \left[ \frac{m}{t} - \frac{n}{t_{s}} \left(1 - \frac{t}{t_{s}}\right)^{n-1}  \right] 
	\end{equation}
	\begin{multline}
	\ddot{a}(t)  = \dot{a}(t) \left[ \frac{m}{t} - \frac{n}{t_{s}} \left(1 - \frac{t}{t_{s}}\right)^{n-1}  \right] + \\ a(t) \left[ - \frac{m}{t^2} + \frac{n(n-1)}{t^2} \left(1 - \frac{t}{t_{s}}\right)^{n-2}  \right]    
	\end{multline}
	where the overdots now denote differentiation with respect to time. From this, one can see that for $1<n<2$ $\dot{a} (0) \to \infty $ and $ \dot{a} (t_{s}) = \frac{m a_{s}}{t_{s}} = $ const. ,  while $a(t_{s}) = a_{s} $ , $\ddot{a}(0) \to \infty $ and $ \ddot{a} (t_{s}) \to -\infty $ and we have sudden future singularities. Furthermore, it was shown in \cite{Dabrowski:2012eb} that for the simplified case of the scale factor (20) with $m=0$, one can get w-singularities for $ n > 0 $ and $ n \neq 1 $. The scale factor for the case $ m = 0 $ takes the form \begin{equation}
	a(t) = a_{s} \exp \left(1 - \frac{t}{t_{s}}\right)^{n}
	\end{equation} and we will be using this form of the scale factor for this section.  With this cleared up, we would firstly like to find expressions the w-parameter in the standard general relativistic cosmology.
	\\
	\\
	The standard Friedmann equation in natural units is \begin{equation}
	H^2 = \frac{\rho}{3}
	\end{equation}
	Considering that the pressure density is related to the energy density as an equation of state of the form (1), we consider the form of $f(\rho) $ as \begin{equation}
	f(\rho) = r^{\alpha}
	\end{equation}
	where $\alpha$ is an arbitrary constant. Considering this form of the EOS, we can write the w-parameter for this cosmology using (18-19) as \begin{equation}
	w = -3^{\alpha-1} \left(\frac{n^2 \left(1-\frac{t}{t_{s}}\right)^{2 (n-1)}}{t_{s}^2}\right)^{\alpha-1}-1
	\end{equation}
	For the w-parameter as expressed above, we have the following observations : \begin{itemize}
		\item For n = 1, no w- singularities occur as is the case in the usual scenario with conventional equation of state. 
		\item For $\alpha < 0 $ , w-singularities occur for all values of positive values of n besides unity but w-singularities do not occur for any negative values of n
		\item For $\alpha > 0 $ we see a very interesting behaviour. In this case, completely in contraction to what happens in the usual case, no w-singularities occur for positive values of n ($n>0$ ) but they occur only when n has negative values ($n<0$). Hence, here we see the first sign of departure in the occurence conditions of w-singularities when one considers inhomogeneous equations of state.  
	\end{itemize}
	Just by incorporating an inhomogeneous EOS in our cosmological setup, we see that one can immediately start to see differences in the occurence conditions of the type V singularities from the usual conditions. The second form of the EOS that we now consider to move further is \begin{equation}
	f(\rho) = \frac{C}{(\rho_{o} - \rho)^{\gamma}}
	\end{equation}
	where C is a positive constant and $\gamma$ is an arbitrary constant. Using this form of the EOS and eqn. (18-19), we get the w-parameter as \begin{equation}
	w = -\frac{C t_{s}^2 \left(1-\frac{t}{t_{s}}\right)^{2-2 n} \left(\rho_{o}-\frac{3 n^2
			\left(1-\frac{t}{t_{s}}\right)^{2 (n-1)}}{t_{s}^2}\right)^{-\gamma}}{3 n^2}-1
	\end{equation}
	For the w-parameter as expressed above, we have the following observations : \begin{itemize}
		\item For n = 1, no w- singularities occur as is the case in the usual scenario with conventional equation of state. 
		\item For $\gamma > 0 $ , w-singularities occur for all values of positive values of n besides unity but w-singularities do not occur for any negative values of n
		\item For $\gamma < 0 $ we again see a very interesting behaviour. In this case, no w-singularities occur for positive values of n ($n>0$ ) but they occur only when n has negative values ($n<0$).
	\end{itemize}
	So we see that when one incorporates an inhomogeneous EOS even in a general relativistic cosmology, then one starts to see some very evident new rules under which w-singularities form and in particular, the telling thing is that now w-singularities can also form for negative values of n.
	\\
	\\
	We now proceed towards the Chern-Simons cosmomlogy for which the Friedmann equation is of the form of (5). Using the form of $f(\rho)$ as described in (20) and the ansatz (18), we arrive at the expression for the w parameter as \begin{equation}
	w = \frac{t_{s}^2 3^{\alpha-1} \left(1-\frac{t}{t_{s}}\right)^{2-2 n} \left(\frac{n^2
			\left(1-\frac{t}{t_{s}}\right)^{2 (n-1)}}{t_{s}^2}-\frac{n^4 \beta \left(1-\frac{t}{t_{s}}\right)^{4
				(n-1)}}{t_{s}^4}\right)^\alpha}{n^2 \left(\frac{n^2 \beta \left(1-\frac{t}{t_{s}}\right)^{2
				(n-1)}}{t_{s}^2}-1\right)}-1
	\end{equation}
	For the w-parameter as expressed above, we have the following observations : \begin{itemize}
		\item For n = 1, no w- singularities occur as is the case in the usual scenario with conventional equation of state. 
		\item For $\alpha > 0 $ , w-singularities occur for all values of positive values of n besides unity but w-singularities do not occur for any negative values of n
		\item For $\alpha < 0 $ we see that no w-singularities occur for positive values of n ($n>0$ ) but they occur only when n has negative values ($n<0$).
	\end{itemize}
	Now we use the form of the EOS as given in (22) and find the w- parameter in this case to be \begin{equation}
	w = -\frac{C \left(\frac{3 n^4 \beta \left(1-\frac{t}{t_{s}}\right)^{4 (n-1)}}{t_{s}^4}-\frac{3 n^2
			\left(1-\frac{t}{t_{s}}\right)^{2 (n-1)}}{t_{s}^2}+\rho_{o} \right)^{-g}}{\frac{3 n^2
			\left(1-\frac{t}{t_{s}}\right)^{2 (n-1)}}{t_{s}^2}-\frac{3 n^4 \beta \left(1-\frac{t}{t_{s}}\right)^{4
				(mn-1)}}{t_{s}^4}}-1
	\end{equation}
	And now we make inferences about w-singularities as evident from the parameter as above to be \begin{itemize}
		\item For n = 1, no w- singularities occur as is the case in the usual scenario with conventional equation of state. 
		\item For $\gamma > 0 $ , w-singularities occur for all values of positive values of n besides unity but w-singularities do not occur for any negative values of n
		\item For $\gamma < 0 $ we see something which we haven't seen in any case till now ; w-singularities occur for both positive and negative values of n (besides unity ) ! 
	\end{itemize}
	So this completes this discussion on the Chern-Simons cosmology. We again see some new conditions for the occurence of type V singularities which we didn't see in the usual case of the standard cosmology, in particular seeing that under the condition that $\gamma < 0$, both positive and negative values of n can give out type V singularities. 
	\\
	\\
	Now we consider the asymptotically safe cosmology which is given by the Friedmann equation (14), for which we firstly use an EOS of the form (20) to get the w-parameter in this case as, \begin{equation}
	w = -\left(\frac{n^4 \left(1-\frac{t}{t_{s}}\right)^{4 (n-1)}}{t_{s}^4 N}\right)^{n-1}-1
	\end{equation}
	For the w-parameter as expressed above, we have the following observations : \begin{itemize}
		\item For n = 1, no w- singularities occur as is the case in the usual scenario with conventional equation of state. 
		\item For $\alpha > 0 $ , w-singularities will occur only for all negative values of n besides unity but w-singularities do not occur for any positive values of n
		\item For $\alpha < 0 $ we see that w-singularities only occur for positive values of n ($n>0$ ) but they do not occur when n has negative values ($n<0$).
	\end{itemize}
	One interesting thing to note here is that one sees w-singularities in this scenario when $\alpha$ and n have different signs, contrary to most of the aforementioned cases when they were occuring in the cases when both these quantities had the same sign. Now we consider the EOS of the form (22), for which the w-parameter has the value \begin{equation}
	w = -\frac{C t_{s}^4 N \left(1-\frac{t}{t_{s}}\right)^{4-4 n} \left(\rho_{o}-\frac{n^4 \left(1-\frac{t}{t_{s}}\right)^{4 (n-1)}}{t_{s}^4 N}\right)^{-\gamma}}{n^4}-1
	\end{equation}
	For this expression of the w-parameter, we can draw the following inferences \begin{itemize}
		\item For n = 1, no w- singularities occur as is the case in the usual scenario with conventional equation of state. 
		\item For $\gamma > 0 $ , w-singularities occur for all values of positive values of n besides unity but w-singularities do not occur for any negative values of n
		\item For $\gamma < 0 $ we see the same behaviour as in the case of the Chern-Simons cosmology which is that w-singularities occur for both positive and negative values of n besides unity.
	\end{itemize}
	For the asymptotically safe cosmology case, we see that similar to the Chern-Simons scenario w-singularities can occur for both positive and negative values here for $\gamma<0$. 
	\\
	\\
	Finally, we now move towards the f(R) gravity cosmology with the Friedmann equation (4), where the w-parameter for the EOS (20) takes the form \begin{equation}
	w = -12^{\alpha-1} \left(\frac{\eta \left(12 n^2 \left(1-\frac{t}{t_{s}}\right)^{2 n}-\eta (t_{s}-t)^2\right)}{11 \eta (t_{s}-t)^2-18 n^2 \left(1-\frac{t}{t_{s}}\right)^{2 n}}\right)^{\alpha-1}-1
	\end{equation} 
	For the w-parameter as expressed above, we have the following observations : \begin{itemize}
		\item For n = 1, contrary to the other cases we have considered till now, one can have a w-singularity but that is possible only in the extreme case that $\alpha \to \infty$ which is not pretty realistic to expect but in principle singularities can appear in this case. 
		\item The most interesting thing that comes out when one considers this scenario is that w-singularities do not occur for any value of n and $\alpha$ ! For both positive and negative values of $\alpha$ and n, the w-parameter remains regular and does not diverges.
	\end{itemize}
	After this, considering the EOS of the form (22) lends us the w-parameter as \begin{equation}
	w = \frac{C \left(\frac{18 n^2 \left(1-\frac{t}{t_{s}}\right)^{2 (n-1)}}{t_{s}^2}-11 \eta\right) \left(\frac{12 \eta \left(\frac{12 n^2 \left(1-\frac{t}{t_{s}}\right)^{2 (n-1)}}{t_{s}^2}-\eta\right)}{\frac{18 n^2
				\left(1-\frac{t}{t_{s}}\right)^{2 (n-1)}}{t_{s}^2}-11 \eta}+\rho_{o}\right)^{-\gamma}}{12 \eta \left(\frac{12 n^2 \left(1-\frac{t}{t_{s}}\right)^{2 (n-1)}}{t_{s}^2}-\eta\right)}-1
	\end{equation}
	For this expression of the w-parameter, we can draw the following inferences \begin{itemize}
		\item For n = 1, no w- singularities occur in this scenario even for extreme values of $\gamma$. 
		\item Similar to the case for the EOS (20), even in this scenario no w-singularities are possible for both positive or negative values of $\gamma$ and n ! 
	\end{itemize}
	From the very interesting discussion on w-singularities in this scenario, we see that such singularities realistically not even form in this f(R) gravity cosmology when one considers inhomogeneous EOS. This is a sparkling realization and just goes to show that when talking about type V singularities, the form of the equation of state influences the creation of these singularities the most, even more important than what the background cosmology is. The fact that the f(R) gravity cosmology is doing such a removal is also in pretty tune with previous work \cite{Trivedi:2022ngt}, where again there was an extended discussion on the occurence conditions of type V singularities and the f(R) model showed the most departure from the conventional occurence conditions. 
	\\
	\\
	\section{Concluding remarks }
	In this paper we have discussed in detail about the formation of type V singularities in various cosmologies after considering inhomogeneous equations of state. We firstly discussed in brief the various types of cosmological singularities being studied heavily in recent times, after which we discussed the various cosmologies we have considered in our work briefly. We then discussed the two forms of inhomogeneous EOS that we have taken into account in our work and discussed their motivations. We then analysed the w-parameter for both these forms of EOS for an exponential form of the scale factor ansatz. From this we were able to see interesting inferences in the standard general relativistic cosmology itself, like w-singularities occuring for negative values of n as well for particular values of the parameters $\alpha$ and $\gamma$ due to the EOS. We then discussed it in the context of more exotic cosmologies in Chern-Simons and an asymptotically safe cosmology for which we again had interesting inferences. But possibly the most interesting one the f(R) gravity cosmology that we considered, where we showed that the f(R) cosmology completely gets rid of type V singularities when one uses either of the two forms of the inhomogenous EOS that we have considered. This work goes to show that the creation and formation of type V singularities is influenced most strongly by the form of the equation of state that one considers, way more so than what background cosmology one chooses. An interesting future endeavour could be based on seeing what kind of endeavours one sees when one does this same kind of analysis with a series ansatz form for the scale factor (of the form discussed in \cite{Fernandez-Jambrina:2010ngm} ) but we will not be pursuing that here.
	\section*{Acknowledgements}
	The author would like to thank Maxim Khlopov, Sergei Odintsov and Tatiana Vulfs for very helpful discussions on various subjects related to this work. The author would also like to thank the referee of this paper for their very constructive comments on the work.

	\bibliography{JSPJMJcitewinhoes.bib}
	
	\bibliographystyle{unsrt}
	
\end{document}